\documentclass[aps,superscriptaddress,twocolumn,longbibliography]{revtex4-1} 
\usepackage{graphicx} \usepackage{color}
\usepackage{setspace} 
\usepackage[margin=1in]{geometry}
\usepackage{amsmath}
\usepackage{amssymb,braket}
\usepackage{ulem}
\usepackage[colorlinks = true]{hyperref}

\usepackage{titlesec}
\begin{document}
\title{Non-Markovian Dephasing and Depolarizing Channels}
\author{U. Shrikant}
\email{shrik@poornaprajna.org}
\affiliation{Poornaprajna Institute of Scientific Research,
Sadashivnagar, Bangalore, India}
\affiliation{Manipal Academy of Higher Education, Manipal, India}
\author{R. Srikanth}
\email{srik@poornaprajna.org}
\affiliation{Poornaprajna Institute of Scientific Research,
Sadashivnagar, Bangalore, India}
\author{Subhashish Banerjee}
\email{subhashish@iitj.ac.in}
\affiliation{Indian Institute of Technology, Jodhpur, Rajasthan, India}

\begin{abstract} 
We  introduce   a  method  to  construct   non-Markovian  variants  of
completely  positive (CP)  dynamical maps,  particularly, qubit  Pauli
channels.   We   identify  non-Markovianity  with  the   breakdown  in
CP-divisibility    of    the    map,    i.e.,    appearance    of    a
not-completely-positive  (NCP) intermediate  map.   In particular,  we
consider  the   case  of  non-Markovian  dephasing   in  detail.   The
eigenvalues of the Choi matrix of  the intermediate map crossover at a
point which corresponds to a  singularity in the canonical decoherence
rate of  the corresponding  master equation, and  thus to  a momentary
non-invertibility of the map.   Thereafter, the rate becomes negative,
indicating non-Markovianity.  We quantify  the non-Markovianity by two
methods, one  based on CP-divisibility  (Hall et al., PRA  89, 042120,
2014), which  doesn't require optimization but  requires normalization
to   handle   the   singularity,   and  another   method,   based   on
distinguishability  (Breuer  et al.   PRL  103,  210401, 2009),  which
requires optimization but is insensitive to the singularity.
\end{abstract}
\maketitle

\section{Introduction}

Quantum technologies  have now  advanced to a  stage where  effects of
memory and its manipulation are expected to play a crucial role in the
theoretical as well  as experimental developments of  the field.  This
necessitates  a   proper  understanding  of   non-Markovian  phenomena
\cite{vega2017dynamics,             RHP14,             li2017concepts,
  vacchini2011markovianity,                     vacchini2012classical,
  breuer2016nonmarkovian, bhattacharya2018resource} in  the context of
open  quantum  systems \cite{grabert1988quantum,  banerjee2000quantum,
  banerjee2003general, breuer2002theory}.

A  classical  (discrete)  stochastic  process $X_t$  $(t  \in  I$)  is
Markovian if the  conditional probability for the  $n$th outcome $x_n$
satisfies:  $P(x_n~|~x_{n-1}; \cdots  ; x_0)  = P(x_n\vert  x_{n-1})$,
i.e., there  is no  memory of  the history  of values  of $X$.   If an
experiment can access only one-point probability vectors, $P(x)$, then
the stochastic  evolution can  be represented  in terms  of transition
matrices connecting  initial and final probability  vectors: $P(x_1) =
\sum_j T(x_1|x_0)  P(x_0)$, where  $T$ has suitable  normalization and
positive  properties.   For  a Markovian  process,  such  ``stochastic
matrices''  compose  according  to  $T(x_k|x_i)  =  \sum_j  T(x_k|x_j)
T(x_j|x_i)$ for any $j$ intermediate between  $k$ and $i < k$. In this
sense, a Markovian process is \textit{divisible}.

A non-Markovian process is not necessarily divisible (because matrices
$T(x_k|x_j)$ may not be well  defined unless $j=0$), instead requiring
the full  hierarchy of  conditional probabilities.   Nevertheless, for
$k>j>0$,  assuming  invertibility  of   $T(x_j|x_0)$,  we  can  define
$T(x_k|x_j)  =  \sum_j  T(x_k|x_0)   T(x_0|x_j)  =  \sum_j  T(x_k|x_0)
T^{-1}(x_j|x_0)$,  though  this  matrix  may not  be  positive.   

The vector $w(x) \equiv q  P_1(x) - (1-q)P_2(x)$ for two distributions
$P_1$ and $P_2$  has the physical signifance that  the minimum failure
probability to  distinguish $P_1$  and $P_2$  in a  single measurement
$p_{\rm  min}^{\rm  fail}  = \frac{1-||w||_1}{2}$,  where  $||v(x)||_1
\equiv \sum_x |v(x)|$ is the $L_1$ norm.  A fundamental result here is
that a classical stochastic process is divisible (read: Markovian) iff
the distinguishability  of two  distributions is  non-increasing under
the process.

It isn't  straighforward to define quantum  non-Markovianity because a
quantum     realization    of     the    conditional     probabilities
$P(x_n|x_{n-1},\cdots,  x_0)$ would  seem to  require conditioning  on
measurement   interventions,   bringing   to  the   fore   issues   of
non-commutativity and  measurement disturbance.  Perhaps, there  is no
unique,   context-independent  definition   of  quantum   Markovianity
\cite{li2017concepts}.   Here, we  use  a  definition of  Markovianity
based  on  divisibility  (in  specific,  \textit{CP-divisibility})  or
distinguishability,    which    needn't    refer    to    measurements
\cite{SB2017nm, kumar2017nonmarkovian}.  In general, these definitions
aren't equivalent in the quantum domain: Markovian \`a la divisibility
implies  Markovian  \`a  la  distinguishability, but  not  vice  versa
\cite{chrusinski2011measures,   liu2013nonunital,  hall2014canonical},
though  they  are  shown  to  be equivalent  for  all  bijective  maps
\cite{bylicka2017constructive}.

CP-divisibity  is   the  requirement   that  the  time   evolution  be
characterized     by     linear,      trace-preserving     CP     maps
$\mathcal{E}_{t_k,t_j}$   ($t_k\ge    t_j   \ge   t_0$)    such   that
$\mathcal{E}_{t_k,t_i}  =  \mathcal{E}_{t_k,t_j}\mathcal{E}_{t_j,t_i}$
for  any   intermediate  time  $t_j$.   Under   quantum  non-Markovian
evolution,   an  intermediate   map  $\mathcal{E}_{t_k,t_j}$   may  be
not-completely-positive (NCP) \cite{jordan2004dynamics}, indicative of
correlations between the system and the environment \cite{RHP10}.

The  lower  bound on  the  probability  of discriminating  two  states
$\rho_1$   and   $\rho_2$  in   one   shot   with  an   optimal   POVM
$\{T,\mathbb{I}-T\}$,  is  known  to  be  $p_{\rm  min}^{\rm  fail}  =
\frac{1-||\Delta||_1}{2}$, where $\Delta \equiv q\rho_1 - (1-q)\rho_2$
is the  Helstrom matrix.  Under  a CP-divisible (identified  here with
Markovian)   process  $p_{\rm   min}^{\rm  fail}$   is  non-decreasing
\cite{kossakowski1972,  ruskai1994}.  Thus,  the  decrease of  $p_{\rm
  min}^{\rm fail}$ (or,  equivalently, increase in distinguishability)
for sometime  indicates non-Markovianity, suggestive of  an underlying
memory  in the  process about  system's initial  state or  information
backflow from  the environment.  The differential  CP-divisible map is
characterized by a time-local  generalization of the Lindblad equation
\cite{gorini1976,   lindblad1976}  with   positive  decoherence   rate
\cite{hall2014canonical}.

Here,  we shall  consider  the problem  of constructing  non-Markovian
versions  of  familiar  Markovian   maps,  in  specific,  qubit  Pauli
channels. An example is the  dephasing channel, wherein a state $\rho$
evolves according to the evolution: 
\begin{equation}
\rho \longrightarrow (1-\kappa)I \rho I + \kappa Z \rho Z.
\label{eq:deph}
\end{equation}
Here,   $\kappa$,   the   ``channel  mixing   parameter'',   increases
monotonically  from  0  (noiseless  case)  to  $\frac{1}{2}$  (maximal
dephasing).     The   operator-sum    representation   of    map   Eq.
(\ref{eq:deph}),  $\rho \rightarrow  \sum_{j=I,Z} K_j  \rho K_j^\dag$,
corresponds to the Kraus operators:
\begin{align}
K_I \equiv \sqrt{1-\kappa}I;\quad\quad
K_Z \equiv \sqrt{\kappa}Z.
\label{eq:DWCM}
\end{align}
Our work  is motivated to  extend this  to the most  general dephasing
channel described by the form:
\begin{align}
K_I(p) &= \sqrt{[1 + \Lambda_I(p)](1-p)} I; \nonumber \\
K_Z(p) &= \sqrt{[1 + \Lambda_Z(p)]p} \; Z,
\label{eq:nmdeph}
\end{align}
and to study the conditions on  $\Lambda_j$ under which the channel is
non-Markovian. This has its roots in the open system dynamics modeling
random    telegraph   noise    \cite{daffer2004depolarizing}.    Here,
$\Lambda_j(p)$ $(j=I,  Z)$ are real  functions and $p$ is  a time-like 
parameter running  monotonically from 0 to  $\frac{1}{2}$.  
By ``time-like'' is meant that $p$ increases monotonically
with time (according to a functional dependence whose details are
not important here.)
We recover
Eq.  (\ref{eq:DWCM})  by setting  $\Lambda_I = \Lambda_Z=0$,  with $p$
effectively becoming $\kappa$.

This work  is arranged as  follows.  In Section  \ref{sec:NMdeph}, the
general  dephasing  channel  in  the form  Eq.   (\ref{eq:nmdeph})  is
derived, and some salient features are noted, among them a singularity
that occurs in  the intermediate map at the crossover  between its two
eigenvalues.   In  Section   \ref{sec:neg},  the  non-Markovianity  is
quantified   using   negative   canonical  decoherence   rate,   which
essentially  measures  how  far  the  instantaneous  intermediate  map
deviates from CPness.   A singularity is encountered  at the crossover
point,  which  is dealt  with  using  a normalization  procedure.   In
Section \ref{sec:sing} we point out  that the singularity represents a
momentary failure  of invertibility  of the  map, but  is nevertheless
harmless.  In Section \ref{sec:TD}, we obtain the trace-distance-based
distinguishability measure  of non-Markovianity. This  measure doesn't
require normalization, and  is shown to be  qualitatively in agreement
with the negative decoherence based measure.  After a brief discussion
of  extending this  method  to non-Markovian  depolarizing in  Section
\ref{sec:depol},  we  conclude  in Section  \ref{sec:conclu},  with  a
discussion  of some  general features  of the  non-Markovian dephasing
channel introduced here.

\section{Non-Markovian dephasing \label{sec:NMdeph}}

The completeness  condition imposed  on Eq.   (\ref{eq:deph}) requires
that:
\begin{align}
 (1-p) \Lambda_I(p) + p  \Lambda_Z(p)  = 0;
\quad\quad 0 \le p \le \frac{1}{2}, 
\end{align} 
implying $ \Lambda_I(p) = - \alpha p $ and $ \Lambda_Z(p) = \alpha (1-
p)    $,    where   $\alpha$    is    real    number.    Then,    from
Eq. (\ref{eq:nmdeph}), we have:
\begin{align}
K_I(t)  &= \sqrt{[1  - \alpha  p](1-p)} \;  I \equiv \sqrt{(1-\kappa)} I \nonumber \\ 
K_Z(t)  &=
\sqrt{[1 + \alpha (1-p)]p} \; Z \equiv \sqrt{\kappa} Z,
\label{eq:nmdephase2}
\end{align} 
which  reduces  to  conventional dephasing  Eq.   (\ref{eq:deph})  for
$\alpha\rightarrow0$.  Here we  choose $0 \le \alpha  \le 1$, ensuring
that the modified dephasing is CP.

Given a quantum map evolving a system  from time 0 to time $t$ through
$s$,    defined    by     the    composition    $\mathcal{E}(t,0)    =
\mathcal{E}(t,s)\mathcal{E}(s,0)$, we can define the intermediate map:
\begin{equation}
\mathcal{E}(t,s) \equiv \mathcal{E}(t,0)\mathcal{E}(s,0)^{-1},
\label{eq:imap}
\end{equation} 
provided  $\mathcal{E}(s,0)$  is  invertible.  This  may  be  computed
directly using matrix inversion \cite{rajagopal2010kraus, usha2012} of
the dynamical map \cite{sudarshan1961stochastic}.

Here  we  derive  it  by  ``vectorizing''  the  density  operator  and
representing the superoperator $\mathcal{E}$ as a corresponding matrix
operation,  using  the  identity  $\widehat{ABC}  =  (C^T  \otimes  A)
\hat{B}$  \cite{RHP10}.  The  intermediate  map is  derived by  matrix
inversion,    and   applied    to    the    vectorized   version    of
$(\ket{00}+\ket{11})$. ``Devectorizing'' this gives the Choi matrix of
the intermediate map.
\begin{equation}
\chi = (\mathcal{E}(t,s)\otimes I)(\ket{00} + \ket{11}).
\end{equation}
By  Choi-Jamiolkowski  isomorphism,  matrix  $\chi$  is  positive  iff
$\mathcal{E}(t,s)$    is     CP    \cite{omkar2013dissipative}.     If
$\mathcal{E}(t,s)$  is   NCP,  then  the  map   $\mathcal{E}(t,0)$  is
non-Markovian.

Consider  an  intermediate  interval   bounded  between  $p^\ast$  and
$p_\ast$, with $0 < p_\ast < p^\ast \le \frac{1}{2}$.  The Choi matrix
for intermediate  map, $\mathcal{E}(\alpha, p^\ast,p_\ast)$,  is found
to be
\begin{align}
M_{\rm Choi} \equiv \left(
\begin{array}{cccc}
1 & 0 & 0 & \frac{(p^\ast-\alpha_-)(p^\ast-\alpha_+)}{(p_\ast-\alpha_-)(p_\ast-\alpha_+)} \\
0 & 0 & 0 & 0 \\
0 & 0 & 0 & 0 \\
\frac{(p^\ast-\alpha_-)(p^\ast-\alpha_+)}{(p_\ast-\alpha_-)(p_\ast-\alpha_+)}
 & 0 & 0 & 1 \\
\end{array}
\right), \label{eqn:choi2}
\end{align}
 where
\begin{align}
\alpha_\pm = \frac{\pm\sqrt{\alpha^2+1}+\alpha+1}{2 \alpha}.
\label{eq:alfa+-}
\end{align} 
The non-vanishing eigenvalues $  \lambda_I$ and $\lambda_Z$ of $M_{\rm
  Choi}$ in Eq. (\ref{eqn:choi2}) are
\begin{align}
\lambda_I(\alpha,p^\ast,p_\ast) 
   & = 1 + \frac{\left(\alpha_--p^\ast\right)\left(\alpha_+-p^\ast\right)}
            {\left(\alpha_--p_\ast\right)\left(\alpha_+-p_\ast\right)};\nonumber\\
\lambda_Z(\alpha,p^\ast,p_\ast) 
   & = 1 - \frac{\left(\alpha_--p^\ast\right)\left(\alpha_+-p^\ast\right)}
            {\left(\alpha_--p_\ast\right)\left(\alpha_+-p_\ast\right)}.
\label{eq:maxeV}
\end{align}
This     leads,     according     to     the     Choi     prescription
\cite{choi1975completely, leung2003choi},  to the Kraus  operators for
the intermediate map:
\begin{align}
K_I^{\rm int}(\alpha, p^\ast, p_\ast) &= \sqrt{\epsilon_I\lambda_I(\alpha,p^\ast, p_\ast)}I,
\nonumber \\
K^{\rm int}_Z(\alpha, p^\ast, p_\ast) &= \sqrt{\epsilon_Z\lambda_Z(\alpha,p^\ast, p_\ast)} Z,
\label{eq:intkraus}
\end{align}
where  $\epsilon_I$  (resp.,  $\epsilon_Z$)  is  $+1$  if  $\lambda_I$
(resp., $\lambda_Z$) is positive and $-1$ otherwise. The corresponding
operator  sum-difference  representation  \cite{omkar2015operator}  of
intermedite evolution  is given by $\rho  \longrightarrow \sum_{j=I,Z}
\epsilon_j   K_j^{\rm   int}\rho   K_j^{{\rm  int}\dag}$.    and   the
completeness  relation  is   $\sum_j  \epsilon_j  K_j^{{\rm  int}\dag}
K_j^{\rm  int}=\mathbb{I}$.   Note  that the  intermediate  map  Kraus
operators also preserve the dephasing form Eq. (\ref{eq:nmdephase2}).

\begin{figure}
\includegraphics[width=7cm]{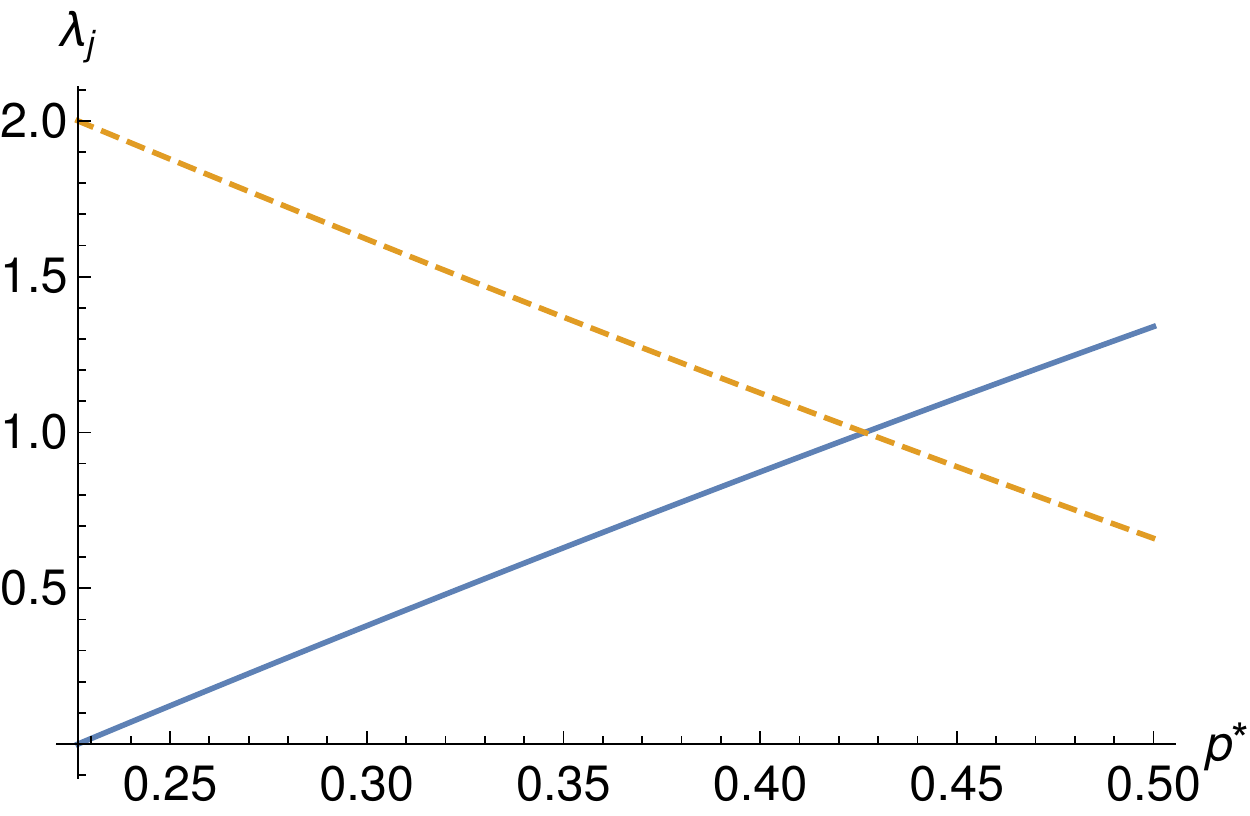}
\caption{(Color online) Eigenvalue $\lambda_I$  (dashed, red line) and
  $\lambda_Z$ (bold, blue  line) for the intermediate  map Choi matrix
  of  the   non-Markovian  dephasing  channel  characterized   by  Eq.
  (\ref{eq:nmdephase2}).   The  intermediate  $p$-range  lies  between
  $p{:=}p_\ast$  and $p{:=}p^\ast$,  where  $p_\ast  < \alpha_-$,  and
  $p^\ast$  is varied  over the  interval $[p_\ast,\frac{1}{2}]$.   At
  $p^\ast {:=} p_\ast$, $\lambda_I=2$ and $\lambda_Z=0$.  At $p^\ast =
  \alpha_-$, the eigenvalues crossover, i.e., $\lambda_I=\lambda_Z=1$,
  and  furthermore  the  channel becomes  maximally  dephasing,  i.e.,
  $\kappa=\frac{1}{2}$ in  Eq.  (\ref{eq:nmdephase2}).   Here, $\alpha
  := 0.3$ and $p_\ast := \alpha_--0.2 \approx 0.23$.}
\label{fig:crossoverdeph}
\end{figure}

\begin{figure}
\includegraphics[width=7cm]{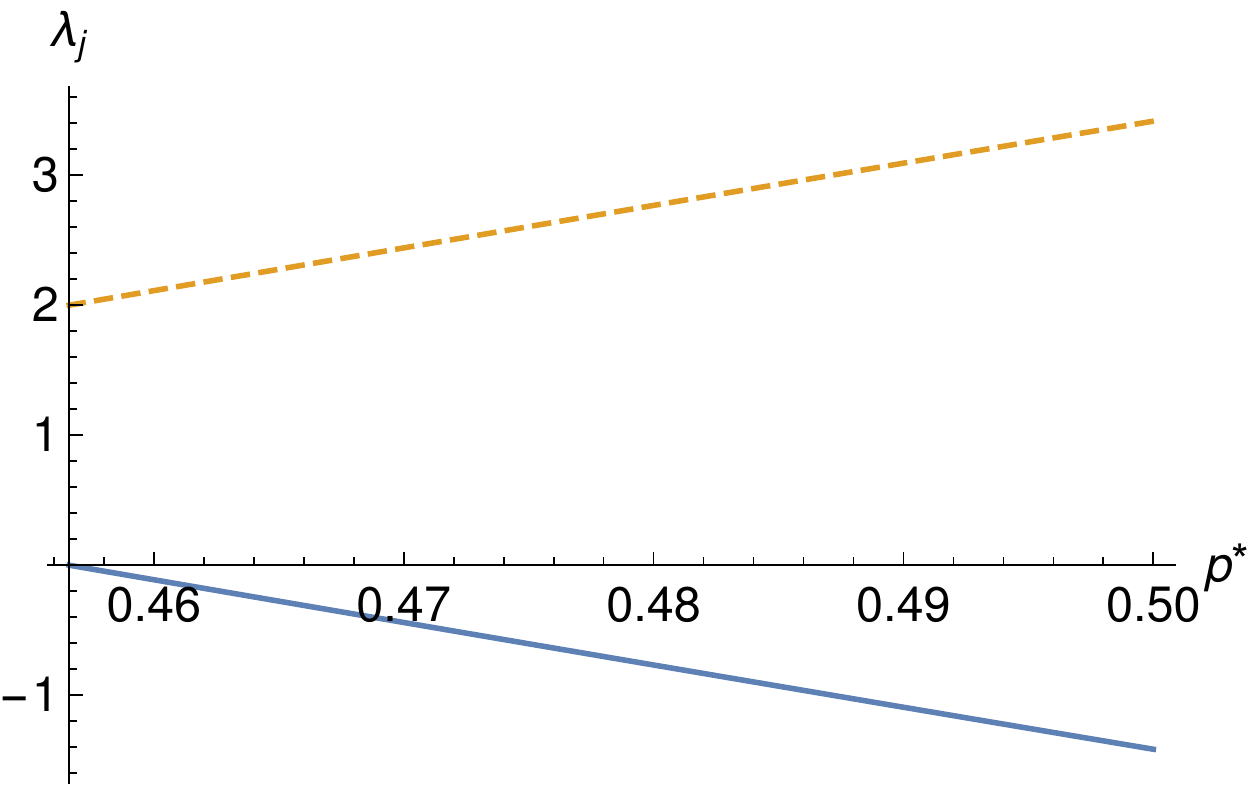}
\caption{(Color online) Eigenvalue $\lambda_I$  (dashed, red line) and
  $\lambda_Z$ (bold, blue  line) for the intermediate  map Choi matrix
  of  the   non-Markovian  dephasing  channel  characterized   by  Eq.
  (\ref{eq:nmdephase2}).  The  intermediate   $p$-range  lies  between
  $p{:=}p_\ast$  and  $p{:=}p^\ast$,  where $\frac{1}{2}  >  p_\ast  >
  \alpha_-$,   and    $p^\ast$   is    varied   over    the   interval
  $[p_\ast,\frac{1}{2}]$.  For $p^\ast > p_\ast$, one finds $\lambda_Z
  <0$.   Thus,  the whole  range  $p^\ast  \in (p_\ast,  \frac{1}{2}]$
corresponds to an  NCP map, demonstrating the  non-Markovianity of the
channel characterized by Eq.   (\ref{eq:nmdephase2}).  Here $\alpha :=
0.3$ and $p_\ast:= \alpha_-+0.03 \approx 0.46$. }
\label{fig:witnegdeph}
\end{figure}

From  Eq. (\ref{eq:maxeV}),  one observes  the following  behavior: if
$p_\ast  <  \alpha_-$   and  $p^\ast$  is  varied   from  $p_\ast$  to
$\frac{1}{2}$, then  the two eigenvalues crossover  at $\alpha_-$ (see
Figure  \ref{fig:crossoverdeph}).  The  crossover  point  is also  the
place where $\kappa=\frac{1}{2}$  in Eq.  (\ref{eq:nmdephase2}), i.e.,
the noise is maximally dephasing.  If $p_\ast > \alpha_-$ and $p^\ast$
is varied from $p_\ast$ to $\frac{1}{2}$, then $\lambda_Z$ is negative
in  the entire  range  $p^\ast \in  (p_\ast,\frac{1}{2}]$ (see  Figure
  \ref{fig:witnegdeph})   and   thus  demonstrates   non-Markovianity.
  Letting   $p^\ast-p_\ast   \rightarrow   0$,  so   that   $\lambda_Z
  \rightarrow 0^-$, we see that  the instantaneous intermediate map is
  NCP here.   This implies that $||M_{\rm  Choi}||_1>1$, and therefore
  the  deviation of  this norm  from 1,  integrated over  the time  of
  evolution, would provide a quantification of non-Markovianity, which
  in fact is the  Rivas-Huelga-Plenio (RHP) measure \cite{RHP10}.  But
  an NCP  intermediate map corresponds to  negative decoherence, which
  suggests a conceptually equivalent, but quantitatively different and
  perhaps computationally simpler method to quantify non-Markovianity,
  based on the integral of the decoherence rate in the master equation
  for    the   negative    rate    period(s).     This   yields    the
  Hall-Cresser-Li-Andersson (HCLA) measure, used here later below.

The  point  $p_\ast=\alpha_-$  represents a  singularity,  since  both
eigenvalues  diverge for  any $p^\ast  \in (p_\ast,\frac{1}{2}]$.   We
  discuss this  matter later  below.  The other  potential singularity
  $p_\ast=\alpha_+$ is not relevant, as the dephasing parameter $p$ is
  assumed  to be  restricted to  the range  $[0,\frac{1}{2}]$, whereas
  $\alpha_+ \in [1,\infty]$.

\section{Negative decoherence rate in the master equation\label{sec:neg}}

\begin{figure}
\includegraphics[width=7cm]{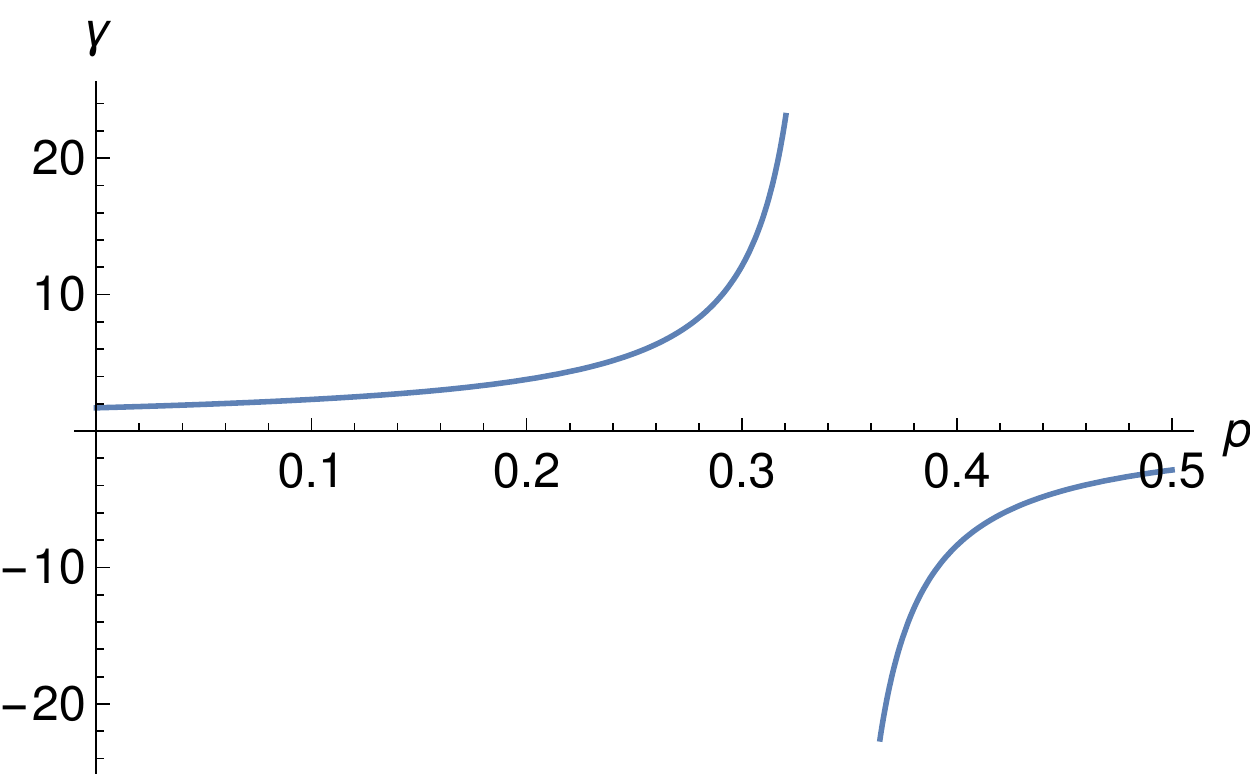}
\caption{Plot of  the decoherence rate  $\gamma$ as a function  of $p$
  for  $\alpha=0.7$.  Note  the singularity  at  $\alpha_-$  $(\approx
  0.34)$, just after which  $\gamma$ becomes negative, indicating that
  the evolution is non-Markovian.}
\label{fig:lindblad}
\end{figure}

The Kraus  representation Eq.  (\ref{eq:nmdephase2}) is  a solution to
the master equation describing dephasing in the canonical form:
\begin{equation}
\frac{d\rho}{dp} = \gamma(p)(-\rho(p) + Z\rho(p) Z).
\label{eq:meq}
\end{equation}
We now  show that the  decoherence rate corresponding  to $\frac{1}{2}
\ge  p  >  \alpha_-$   is  negative,  indicative  of  non-Markovianity
\cite{hall2014canonical}.    By  direct   substitution,  and   letting
$G\equiv1-2\kappa(p)$, one finds:
\begin{align}
\gamma(p) = -\frac{1}{2 G}\frac{dG}{dp} = 
\frac{\frac{1}{2}(\alpha_++\alpha_-)- p}{(p-\alpha_-)(p-\alpha_+)},
\label{eq:lindblad}
\end{align}
from which one sees that the evolution for $p < \alpha_-$ is Markovian
($\gamma\ge0$),   but    becomes   non-Markovian    ($\gamma<0$)   for
$p>\alpha_-$.  The  point $\alpha_-$  itself represents  a singularity
(see Figure \ref{fig:lindblad}).

\begin{figure}
\includegraphics[width=7cm]{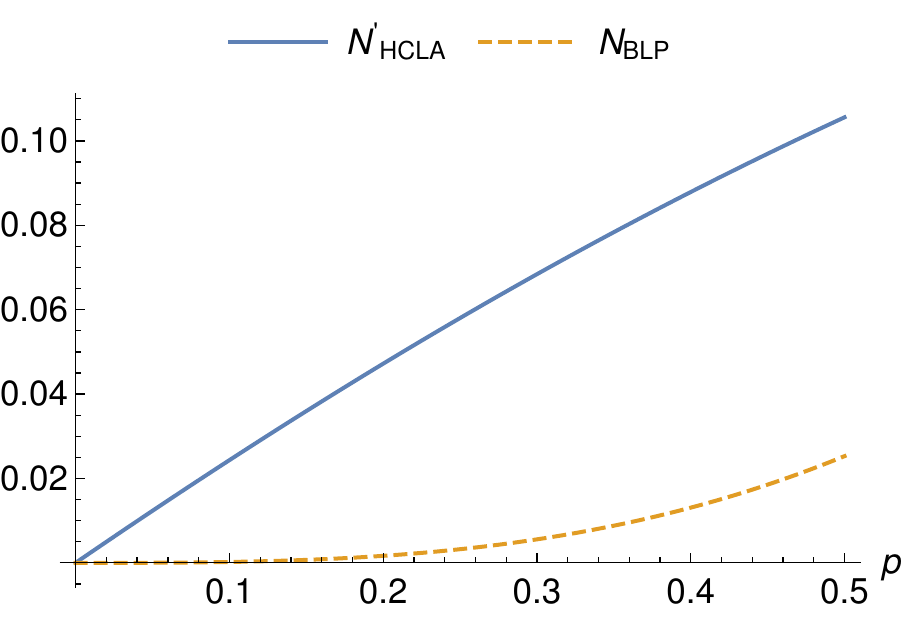}
\caption{(Color  online)  Plot  of  the  normalized  HCLA  coefficient
  $N_{\rm HCLA}^\prime$  (bold, blue  line, Eq.   (\ref{eq:HCLA})) and
  the   BLP  coefficient   $N_{\rm  BLP}$   (dashed,  red   line,  Eq.
  (\ref{eq:NBLP})),  as  a  function of  the  non-Markovian  parameter
  $\alpha$.}
\label{fig:HCLA}
\end{figure}

Following \cite{hall2014canonical}, we want  to quantify the amount of
non-Markovianity  by   $N_{\rm  HCLA}   \equiv  -\int_{\alpha_-}^{1/2}
\gamma(p)dp$, which however, would  diverge because of the singularity
at   $\alpha_-$.   One   remedy,   following  an   idea  proposed   in
\cite{RHP14}, is to replace $-\gamma(p)$ by its normalized version
\begin{align}
\gamma^\prime \equiv \frac{-\gamma}{1-\gamma}= 
\frac{\alpha -2 \alpha  p+1}{\alpha -2 \alpha  p^2+2 p},
\end{align}
from which we can define a normalized HCLA measure:
\begin{align}
N_{\rm  HCLA}^\prime   &\equiv
\int_{\alpha_-}^{1/2}  \gamma^\prime(p)dp \nonumber \\
&= \bigg[\frac{1}{2} \log \left(\alpha+2p-2 \alpha p^2\right) \nonumber\\
&~~-\frac{\alpha \tan ^{-1}\left(\frac{2 \alpha p-1}{\sqrt{-2 \alpha^2-1}}\right)}{\sqrt{-2 \alpha^2-1}}\bigg]_{\alpha_-}^{1/2}
\label{eq:HCLA}
\end{align}
A  plot  of $N_{\rm  HCLA}^\prime$  (bold  line)  is given  in  Figure
\ref{fig:HCLA}.  The monotonic increase  of this measure with $\alpha$
justifies its  being regarded as a  non-Markovianity parameter.  These
results are  directly related  to the RHP  measure, $N_{\rm  RHP}$, of
non-Markovianity \cite{RHP10}, since $N_{\rm HCLA} = \frac{d}{2}N_{\rm
  RHP}$, where $d$ is system dimension \cite{hall2014canonical}, which
here is 2.

\section{The singularity isn't pathological \label{sec:sing}}

The possible non-invertibility  of the time evolution  is discussed in
\cite{andersson2007finding,  RHP14},  in   particular,  the  issue  of
general  consistency conditions  on  such  a map  to  derive a  master
equation, and  the problem of quantification  of non-Markovianity.  In
the present  case, the  singularity at  $p=\alpha_-$ corresponds  to a
time    where    the    trajectories    of    all    initial    states
$\cos(\frac{\theta}{2})\ket0  +  e^{i\phi}\sin(\frac{\theta}{2})\ket1$
differing only  by the azimuthal angle  $\phi$, momentarily intersect.
This  is  because  the   point  $p=\alpha_-$  corresponds  to  maximal
dephasing,   under    which   any   initial   qubit    state   $\rho_1
\equiv  \begin{pmatrix}  a &  b  \\  b^\ast  & 1-a  \end{pmatrix}$  is
transformed  to  $\rho_2   \equiv  \begin{pmatrix}  a  &  0   \\  0  &
  1-a \end{pmatrix}$.  In  other words, all off-diagonal  terms in the
computational  basis  are  killed  off,  making  the  map  momentarily
non-invertible.  Nevertheless, the  singularity isn't pathological, in
the sense  that the density  operator, and consequently the  full map,
are well defined, and invertibility is subsequently recovered.

At   time    $\alpha_-$,   the   intermediate   dynamical    map   Eq.
(\ref{eq:intkraus})  advancing  the state  by  a  small time  interval
$\epsilon$, is acting  on a density operator of the  type $\rho_2$ and
induces the intermediate evolution:
\begin{align}
\rho_2 &\longrightarrow \frac{(1 + Y)}{2}\rho_2 + 
\frac{(1-Y)}{2}Z\rho_2 Z \nonumber \\
    &= \frac{(1+Y)}{2}\rho_2 + \frac{(1-Y)}{2}\rho_2 = \rho_2,
\end{align}
where        $Y       \equiv        \frac{\left(\alpha_--p^\ast\right)
  \left(\alpha_+-p^\ast\right)}          {\left(\alpha_--p_\ast\right)
  \left(\alpha_+-p_\ast\right)}$  is  the  divergent  summand  in  the
expression for $K_I^{\rm int}$ in  Eq.  (\ref{eq:intkraus}) and we set
$p_\ast {:=} \alpha_-$.  Since the singularity in the intermediate map
occurs at the point of maximal dephasing, the infinite term $Y$ has no
effect,  as it  would only  multiply  with off-diagonal  terms in  the
density operator, which vanish.

Similarly,  in  the  master   equation  (\ref{eq:meq})  for  the  rate
$\frac{d\rho}{dp}$, we note that the  divergence of $\gamma(p)$ at the
singularity is rendered  harmless by virtue of the fact  that the term
$\rho(\alpha_-) - Z\rho(\alpha_-)Z$, which it multiplies, vanishes for
the above reason.

\section{Quantifying non-Markovianity via trace distance
\label{sec:TD}}

There are a host of  measures to witness or quantify non-Markovianity,
such as  trace distance,  fidelity, quantum relative  entropy, quantum
Fisher  information,  capacitance  measures; as  well  as  correlation
measures such as mutual information, entanglement, and discord, all of
which are non-increasing under CP-divisible maps, and can thus be used
to witness non-Markovianity \cite{RHP10}.

Here,   we   consider   evolution   of   the   trace   distance   (TD)
\cite{breuer2009measure},  applied  to  the pair  of  initial  states:
$\ket{\psi_0}  = \cos  (\theta/2) \ket{0}+  e^{i \phi}  \sin(\theta/2)
\ket{1}$  and  $\ket{\psi_1}  = -  \sin(\theta/2)  \ket{0}+  e^{i\phi}
\cos(\theta/2) \ket{1}$. For this pair:
\begin{align}
{\rm TD}(\theta,\phi,\alpha, p) &\equiv \frac{1}{2}
{\rm tr}\sqrt{(\rho_0 - \rho_1)^2} \nonumber\\
  &= \bigg[1 - 4 \alpha^2(1-p) p 
(\alpha_+ + \alpha_- - p) \times \nonumber\\
&\quad\quad (2\alpha_+\alpha_-- p)\sin ^2(\theta )  \bigg]^{1/2},
\label{eq:TDgen}
\end{align}
where    $\rho_j    =   \mathcal{E}(\ket{\psi_j}\bra{\psi_j})$,    and
$\mathcal{E}$ represents  the time  evolution under  our non-Markovian
dephasing.  The expression  is independent  of $\phi$,  reflecting the
azimuthal symmetry of the dephasing action \cite{banerjee2007dynamics,
  banerjee2008geometric}.  For $\theta$ where $0  < \theta < 2\pi$, it
may be seen that TD attains a minimum of $\cos(\theta)$ at $\alpha_-$.
The subsequent ($p>\alpha_-$) rise in TD signals non-Markovianity.

\begin{figure}
\includegraphics[width=7cm]{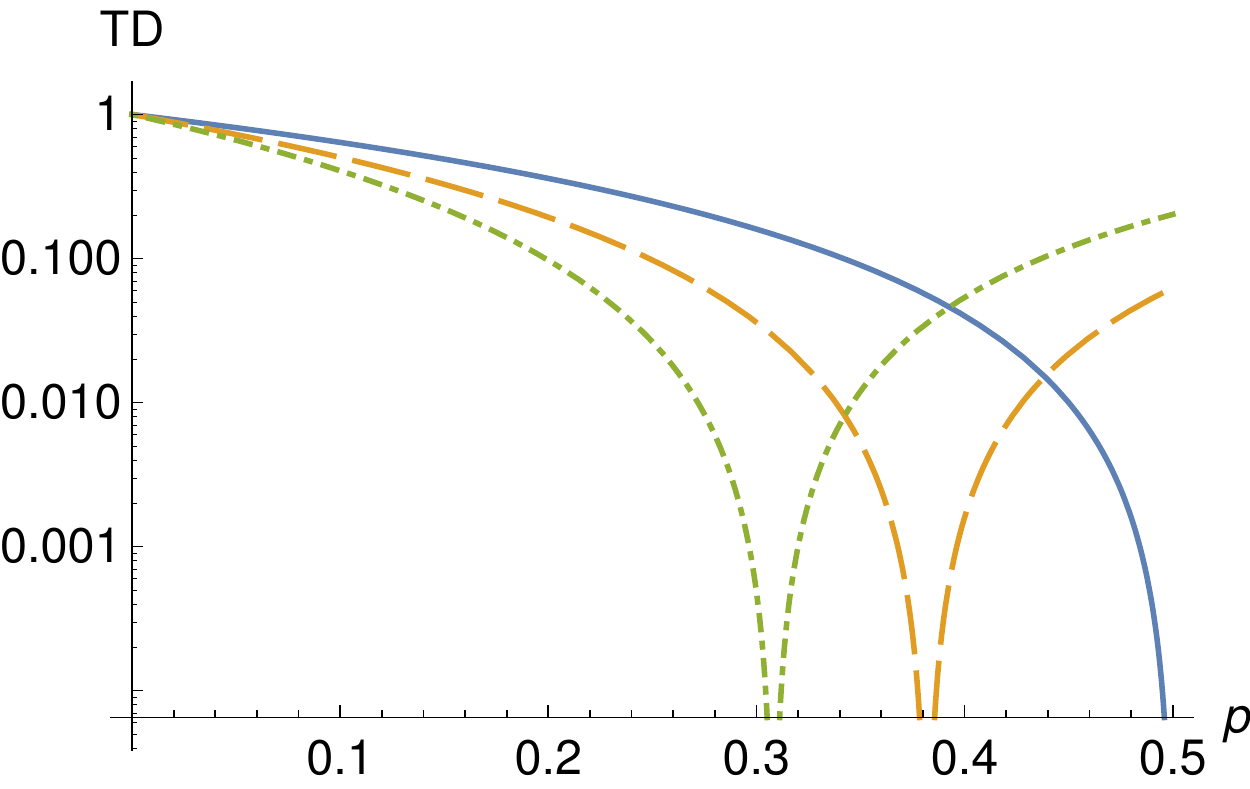}
\caption{Log  plot   of  trace   distance  TD   between  $\rho_0\equiv
  \mathcal{E}(\ket{\psi_0}\bra{\psi_0})$       and       $\rho_1\equiv
  \mathcal{E}(\ket{\psi_1}\bra{\psi_1})$ as a function of $p$
                          with
  $\theta{:=}\frac{\pi}{2}$,   under   the  considered   non-Markovian
  dephasing  noise.   The  bold   (blue)  curve  represents  Markovian
  dephasing, and shows no  recurrence.  The dashed (red, $\alpha=0.5)$
  and     dot-dashed    (green,     $\alpha=0.9$)    show     enhanced
  distinguishability   beyond   their   respective   crossover   point
  $\alpha_-$,  indicative  of   non-Markovianity.   Note  that  larger
  $\alpha$  shows  a  larger  enhancement  region,  suggesting  larger
  non-Markovianity in the sense of BLP \cite{breuer2009measure}.}
\label{fig:TDdeph}
\end{figure}

This pattern  is manifest in  the case of  $\theta=\frac{\pi}{2}$, for
which Eq.  (\ref{eq:TDgen}) reduces to the particularly simple form
\begin{align}
{\rm TD}_{\pi/2}(p)= 2\alpha\left(p-\alpha_-\right)
\left(p-\alpha_+\right).
\label{eq:TD}
\end{align}
This is depicted in  Figure \ref{fig:TDdeph} for various non-Markovian
parameters  $\alpha$.  We  note  in this  Figure  that the  recurrence
region  $(\alpha_-,\frac{1}{2}]$   is  larger  for   larger  $\alpha$,
  suggestive of greater non-Markovianity for larger $\alpha$. 

The BLP measure  of non-Markovianity, denoted $N_{\rm  BLP}$, is given
by:
\begin{align}
N_{\rm BLP}  &= \max_{(\psi_0,  \psi_1)} \int_{\alpha_-}^{1/2} \frac{d{\rm  TD}}{dp} dp
\nonumber \\ &=  \max_\theta 
\bigg[\sqrt{1+ (-1+\frac{\alpha^2}{4})\sin^2(\theta)}-\cos(\theta)
\bigg] \nonumber\\ 
&= \frac{\alpha}{2}.
\label{eq:NBLP}
\end{align}
The  result   is  depicted  as   the  dashed  (red)  line   in  Figure
\ref{fig:HCLA}, and shows  that there is a general  agreement with the
quantification of  non-Markovianity according  to the  normalized HCLA
measure $N_{\rm HCLA}^\prime$.

Here,  following \cite{breuer2009measure},  we have  assumed that  the
pair of states parametrized by  $(\theta, \phi)$, is orthogonal.  This
is   appropriate,   to   enhance  the   contrast   that   demonstrates
non-Markovianity.    Specifically,  note   that  the   TD  in   Figure
\ref{fig:TDdeph}  varies  in  the  range between  1  (initial)  and  0
(maximal dephasing).   If, on  the other hand,  the two  initial stats
were (say)  $\ket0$ and  $\frac{1}{\sqrt{2}}(\ket0 + \ket1)$,  then TD
varies in the smaller range between $\frac{1}{\sqrt{2}}$ (initial) and
$\frac{1}{2}$ (maximal dephasing).

\section{Non-Markovian Depolarizing \label{sec:depol}}

The  depolarizing channel  of a  qubit  transforms state  $\rho$ to  a
mixture of  itself and the  maximally mixed state.   The non-Markovian
version of  the depolarizing  channel can  also be  found in  a manner
analogous to the dephasing channel, which is now discussed briefly.

A  Kraus representation  for the  depolarizing channel would be
$\rho  \longrightarrow  \sum_j  K_j   \rho  K_j^\dag$,  where  $K_I  =
\sqrt{1-p}I$, $K_X = \sqrt{\frac{p}{3}}X$, $K_Y = \sqrt{\frac{p}{3}}Y$
and $K_Z = \sqrt{\frac{p}{3}}Z$.   A potential non-Markovian extension
for them would be
\begin{align}
K_I &= \sqrt{(1+\Lambda_1)(1-p)} \; \quad 
K_X = \sqrt{(1 + \Lambda_2) \frac{p}{3}} X \nonumber\\
K_Y &= \sqrt{(1 + \Lambda_2) \frac{p}{3}} Y; \quad
K_Z = \sqrt{(1 + \Lambda_2) \frac{p}{3}} Z 
\label{eq:nmdepol}
\end{align}
where $\Lambda_k$ ($k  \in \{1,2\}$) is a real function,  and $p$ is a
timelike parameter  that rises monotonically from  0 to $\frac{1}{2}$.
The variables $\Lambda_j$ satisfy the following condition, 
\begin{align}  (1-p) \, \Lambda_1 + p
  \, \Lambda_2 = 0
\label{eq:complete}
\end{align}
as a consequence of the completeness requirement.

In  agreement with  Eq.   (\ref{eq:complete}), we  make the  following
choices: $ \Lambda_1 = - 3 \alpha  p$ and $ \Lambda_2 = 3 \alpha (1-p)
$, where  $\alpha$ is  real. Then,  the non-Markovian  Kraus operators
take the form
\begin{align}
K_I(p)  &= \sqrt{[1  - 3\alpha  p](1-p)} \;  I \nonumber  \\ 
K_X(p)  &= \sqrt{[1 + 3\alpha (1-p)]\frac{p}{3}}X, \nonumber\\
K_Y(p)  &= \sqrt{[1 + 3\alpha (1-p)]\frac{p}{3}}Y, \nonumber\\
K_Z(p)  &= \sqrt{[1 + 3\alpha (1-p)]\frac{p}{3}}Z, \nonumber\\
\label{eq:nmdepol2}
\end{align} 
As  before,   parameter  $\alpha$  may   be  seen  to   represent  the
non-Markovian behavior of the channel, such that setting $\alpha := 0$
reduces the  Kraus operators in  Eq.  (\ref{eq:nmdepol2}) to  those in
the conventional Markovian depolarization channel.
 
\section{Conclusions and discussion \label{sec:conclu}}

We  introduced  a  method   to  construct  non-Markovian  variants  of
completely  positive (CP)  dynamical maps,  particularly, qubit  Pauli
channels,   with   non-Markovianity    defined   by   departure   from
CP-divisibility.    Specifically,    a   one-parameter   non-Markovian
dephasing channel was  studied in detail, which is  characterized by a
singularity in  the canonical decoherence rate  $\gamma$, which occurs
at the crossover  point $\alpha_-$ associated with  the eigenvalues of
the  intermediate  map,  and  where   phase  noise  is  maximal.   The
decoherence    rate    $\gamma$    is     negative    for    $p    \in
(\alpha_-,\frac{1}{2}]$,  indicating  non-Markovianity.   Intuitively,
  this can  be understood as  due to $\kappa$ in  Eq.  (\ref{eq:deph})
  exceeding $\frac{1}{2}$, thereby enhancing distinguishability.

More  precisely,  substituting  the  form  Eq.   (\ref{eq:deph})  into
Eq. (\ref{eq:meq}), one finds that:
\begin{equation}
\gamma = \frac{d\kappa/dp}{1-2\kappa},
\label{eq:gamma}
\end{equation}
which relates the ``channel  mixing rate'' $\frac{d\kappa}{dp}$ to the
decoherence rate $\gamma$.  From Eq. (\ref{eq:gamma}), it follows that
\begin{equation}
\gamma < 0 {\rm~iff~} 
\bigg\{ 
\begin{array}{ll}
\frac{d\kappa}{dp} < 0, & {\rm ~in~case~of~} \kappa < \frac{1}{2} \\
\frac{d\kappa}{dp} > 0, & {\rm ~in~case~of~} \kappa > \frac{1}{2},
\end{array}
\label{eq:caseof}
\end{equation}
with $\kappa=\frac{1}{2}$ representing a  singularity.  In the form of
noise we consider, the second  case in Eq.  (\ref{eq:caseof}) explains
the origin of non-Markovianity.  The  reason is that the derivative of
the ``channel  mixing parameter''  $\kappa$ is always  positive, i.e.,
$\frac{d\kappa}{dp} > 0$.  Thus, $\kappa(p)$ must exceed $\frac{1}{2}$
for non-Markovianity to occur.  In view of Eq.  (\ref{eq:gamma}), this
entails    that   a    singularity    must    be   encountered    when
$\kappa=\frac{1}{2}$, which happens in our case at $p=\alpha_-$.

\begin{figure}[ht]
\includegraphics[width=7cm]{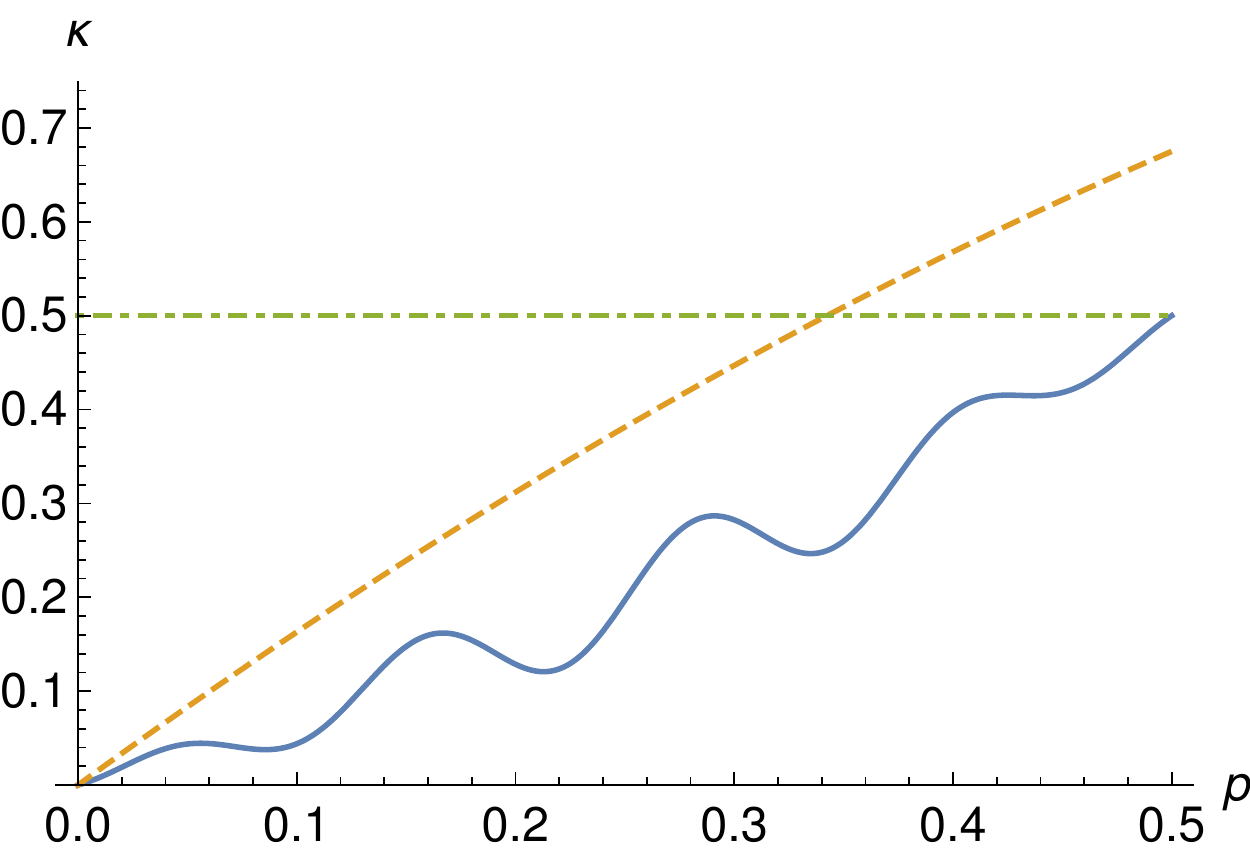}
\caption{Plot of $\kappa(p)$  in Eq.  (\ref{eq:rtn}) as  a function of
  $p$, with $\eta=\frac{1}{2}$ and $\omega=50$ (bold, blue line). This
  dephasing   channel   corresponds  to   the   first   case  of   Eq.
  (\ref{eq:caseof})  and  non-Markovianity   arises  from  regions  of
  negative slope  in the plot.   The dashed (red) line  corresponds to
  the   non-Markovian   dephasing  Eq.    (\ref{eq:nmdephase2})   with
  $\alpha=0.7$. The  mixing rate  $d\kappa/dp$ is never  negative, and
  non-Markovianity     pertains    to     the     first    case     in
  Eq. (\ref{eq:caseof}).}
\label{fig:toy}
\end{figure}

This is illustrated by the  dashed (red) plot in Figure \ref{fig:toy},
which represents  our non-Markovian  dephasing with  $\alpha=0.7$, for
which $\frac{d\kappa}{dp}>0$  throughout the  range $[0,\frac{1}{2}]$.
The point  $\alpha_-$, where  this intercepts  the horizontal  line of
$\kappa=\frac{1}{2}$, is the singularity.  Non-Markovianity comes from
the  positive   mixing  rate  ($\frac{d\kappa}{dp}>0$)  region   $p  >
\alpha_-$.

On  the  other  hand,  non-Markovian dephasing  noise  where  $\kappa$
remains within $[0,\frac{1}{2}]$ as $p$ increases monotonically from 0
to   $\frac{1}{2}$,   corresponds   to   the   first   case   in   Eq.
(\ref{eq:caseof}).    Here,  the   channel   mixing  parameter   can't
monotonically   rise,    i.e.,   there    must   be    regions   where
$\frac{d\kappa}{dp}<0$.  As a simple instance, consider:
\begin{equation}
\kappa(p)  =   p\frac{\left(1  +  \eta   \sin(\omega  p)(1-2p)\right)}
      {\left(1 + \eta(1-2p)\right)},
\label{eq:rtn}
\end{equation}
with $0 \le p \le \frac{1}{2}$, where $\eta$ and $\omega$ are positive
constants characterizing the strength and frequency of the channel.
Such a noisy channel encounters  no singularity, and the non-Markovian
contributions  come   from  the   regions  of  negative   mixing  rate
$\frac{d\kappa}{dp}$, which  arises because  of the sine  function.  A
plot of $\kappa(p)$ for $\eta=\frac{1}{2}$ and $\omega=50$ is the bold
(blue)  plot in  Figure  \ref{fig:toy}. 

We  discussed two  methods  of quantifying  the non-Markovianity,  one
based  on  CP-divisibility  and another  on  distinguishability.   The
former  is derived  from  the  HCLA measure  \cite{hall2014canonical},
based on negative decoherence rates  in the canonical master equation.
This  doesn't require  optimization but  is marked  by a  singularity,
which we  have handled by  using a suitable normalization.   The other
measure is  the BLP  measure \cite{breuer2009measure},  which requires
optimization but is unaffected by the singularity.

Our  method to  construct  a non-Markovian  variant  of the  dephasing
channel  can be  straightforwardly extended  to other  Pauli channels,
e.g.,  bit  flip  or  depolarizing  channels.   Details  such  as  the
level-crossing  feature  of the  eigenvalues  of  the Choi  matrix  of
intermediate map  and the  occurence of  singularities, may  vary from
case to case, presenting new insights.

\acknowledgments

We  are thankful  to  Michael  Hall and  Howard  Wiseman for  fruitful
discussions,  that helped  improve this  manuscript.  SB  acknowledges
support by the project number  03(1369)/16/EMR-II funded by Council of
Scientific and Industrial Research, New  Delhi, India. US and RS thank
DST-SERB, Govt.  of India, for  financial support provided through the
project EMR/2016/004019.
 
\bibliography{QW_NM}

\end{document}